\documentclass{osa-article}

\journal{osajournal}

\articletype{Research Article}

\usepackage{amsmath}
\usepackage{mathrsfs}
\usepackage{graphicx}
\usepackage{siunitx}
\usepackage{enumerate}
\usepackage{mathtools}
\usepackage{mathrsfs}

\def\bra#1{\mathinner{\langle{#1}|}}
\def\ket#1{\mathinner{|{#1}\rangle}}
\def\braket#1{\mathinner{\langle{#1}\rangle}}

\def\bravert{\egroup\,\vrule\,\bgroup}

\newcommand{\vac}{\ket{\text{vac}}}

\newcommand{\sinc}{\text{sinc}}

\begin{document}

\title{Quantum state preparation and one qubit logic from third-order nonlinear interactions}	

\author{Francisco A. Dom\'inguez-Serna,\authormark{1,*} and Karina Garay-Palmett\authormark{2}}

\address{\authormark{1}C\'atedras CONACYT, Centro de Investigaci\'on Cient\'ifica y de Educaci\'on Superior de Ensenada, Apartado Postal 2732, BC 22860 Ensenada, M\'exico\\
\authormark{2}Departamento de \'Optica, Centro de Investigaci\'on Cient\'ifica y de Educaci\'on Superior de Ensenada, Apartado Postal 2732, BC 22860 Ensenada, M\'exico}

\email{\authormark{*}fadomin@cicese.mx} 



\begin{abstract}
We present a study on preparing and manipulating path-like temporal-mode (TM) qubits based on third-order nonlinear interactions. Specifically, we consider the process of frequency conversion via difference frequency generation. To prepare a qubit, we aim to use Gaussian input states to a nonlinear waveguide. The coupling between the input state and a specific TM is maximized, obtaining qubits prepared with fidelities close to one. TMs evolve linearly within the medium; therefore, it is possible to define rotations around any axis contained in the $xy$ plane, allowing spanning the full Bloch sphere in two steps. Particularly, we present a method to obtain any of the Pauli quantum gates by varying geometric or user-accessible parameters in a given experimental configuration. Our study allows for experimentally feasible proposals capable of controllable arbitrary qubit transformations.
\end{abstract}

\section{Introduction \label{sec:intro}}

Quantum information science (QIS) offers advantages in transmission, processing, and information storage that have no classical counterpart \cite{Bravyi2018}. The advances in quantum technologies that allow the manipulation and control of systems at the quantum level are necessary in the development of experimentally feasible applications so as to exploit the promising advantages of QIS. In this regard, many frameworks have been proposed and explored to implement QIS applications, such as trapped ions, nuclear magnetic resonance, photons, superconducting materials, among many others \cite{Hughes2007,Knill01a,Lukens2017,Obrien2010,Neergaard-Nielsen:2010uq,Linke2017,DiVincenzo2000,Childress2013,Wang2016}. Even though there is no consensus on a definitive platform for QIS implementations, it is a widespread belief that photons will play a central role in the ultimate scheme \cite{Obrien2010}, mainly because of their low decoherence sensitivity. The main advantage of photons is also their downside, as they present low to null interaction between them and other parties \cite{Lukens2017}. Fortunately, there are many proposals to overcome this limitation through hybrid wave-particle states \cite{Zhang2020}, which brings more attention to light states and the feasibility of their applications. 

Photon temporal-modes (TM) are orthogonal wave-packets of the electromagnetic field that fully overlap in time, polarization and spatial mode \cite{Smith2007,Reddy2017a,Raymer2019}, soon after that, an interesting application of them was presented as the foundation of quantum pulse gate (QPG) \cite{Brecht2011}. Those works led to a  solid QIS application proposal \cite{Raymer2015}, where the TM structure of photon pairs generated by spontaneous parametric down-conversion (SPDC) was used to form qudits in the TM basis, with other works derived ever since \cite{Qin2015,Morin2019,Reddy2017,Reddy2017a}. SPDC is a flavor of three-wave mixing that manifests in a medium with second-order nonlinearity. In the work in Ref. \cite{Brecht2011}, difference frequency generation (DFG) provides the time evolution that naturally served as unitary operations for the qudit TM-basis. Furthermore, Ansari \textit{et al.} have recently experimentally proven the feasibility of TM qudits for QIS applications on remote state preparation protocols by means of a pulse shaper (PS). \cite{Ansari2020a}. The PS, allows for the preparation of an intense pulse with a specific temporal structure, which together with the second-order nonlinear interaction becomes a TM projector. PSs are at the core-level of the proposed framework for quantum information  \cite{Raymer2015}.


In this work, we study the third-order difference frequency generation (TODFG) process in a nonlinear (NL) material as a means to prepare path-like temporal-mode qubits on demand. The TODFG process in a NL medium involves four fields with frequencies $\omega_1, \omega_2, \omega_3$ and $\omega_4$, that obey the energy conservation $\omega_1-\omega_2 = \omega_3-\omega_4$ and phase-matching. In the undepleted pump approximation with two pump fields in modes 1 and 2, TODFG corresponds to the frequency conversion process between modes 3 and 4, assuming phase-matching conditions are met, i.e., it will be governed by a Hamiltonian of the type $ H \propto \xi \hat{a}_1^\dagger \hat{a}_2 \hat{a}_3^\dagger \hat{a}_4   + h.c. $, where if the first two modes are replaced by the classical amplitudes as $\hat{a}_i \to \alpha_i e^{i\theta_i}$ , we obtain $H \propto \alpha_1 \alpha_2 e ^{i(\theta_1 - \theta_2)} \hat{a}_3^\dagger \hat{a}_4   + h.c.$. It can be seen from this toy model that the process is equivalent to a beam-splitter (BS) of TMs, which has been previously studied in references \cite{Reddy2017a,Brecht2011}. We apply the TM formalism of parametric processes to TODFG, which allows to define qubit preparations and rotations based on external user-accessible parameters.

The paper is organized as follows: in section \ref{sec:model}, the model for the nonlinear process under study is revisited in a comprehensive way, and the computational basis is introduced, as well as its transformations. In section \ref{sec:qubpreprot}, the alternatives to prepare and manipulate qubits for this proposal are presented. In section \ref{sec:specificMAP}, a specific and realistic situation is analyzed, where a fidelity measure of the qubit preparation and rotation is introduced. Finally, conclusions and comments on future work are included. 
 
\section{Model}\label{sec:model} 

We consider the TODFG process confined in a single-spatial mode waveguide, with copropagating and co-polarized fields. In such conditions, the Hamiltonian of the interaction can be written as follows

\begin{equation}
\begin{aligned}
\hat{H}(t) &=  \int dk_{1} dk_{2} dk_3 dk_4 S (k_1,k_2,k_3,k_4) e^{i\Delta \omega t} \\
&\quad \times \hat{a}_1^\dagger (k_1) \hat{a}_2 (k_2) \hat{a}_3^\dagger (k_3) \hat{a}_4 (k_4) + h.c.,
\end{aligned}
\end{equation}
where $\hat{a}^\dagger_1(k_1)$($\hat{a}_2(k_2)$)  is the creation (destruction) operator of pump photons with wave-vector $k_1$ ($k_2$) and $\hat{a}^\dagger_3(k_3)$($\hat{a}_4(k_4)$) is the creation (destruction) operator of signal photons with wave-vectors $k_3$ ($k_4$). In our analysis, we assume that pump fields are highly populated, whilst the signal fields are considered at the single photon level. $\Delta \omega = \omega_1 -\omega_2 +\omega_3 -\omega_4  $, takes into account the energy conservation, and $S (k_1,k_2,k_3,k_4)$ contains the nonlinear electric susceptibility, geometric characteristics of the medium, spatial profiles of the involved modes, and field dispersion dependencies, in a similar fashion as in ref. \cite{Dominguez-Serna2020}. The pump operators are changed for classical amplitudes as  $\hat{a}_i \to \alpha_i (\omega_i) e^{i\theta_i}$, where $\alpha_i (\omega_i) $ takes into account the spectral distribution of the pump fields, such that $\int d\omega |\alpha(\omega_i)|^2 =\braket{\hat{n}_i} $, where $\braket{\hat{n}_i}$ represents the average photon number of the $i$-field per unit time. The pump fields are also assumed of pulsed nature throughout this manuscript, therefore $\braket{\hat{n}_i}$ is proportional to the laser repetition rate $R$, the $i$-th pump power average $P_i$ and inversely proportional to its bandwidth $\sigma_i$ when a Gaussian envelope is considered. After integration over the spatial coordinates and $k_1$ and $k_2$, the interaction Hamiltonian is rewritten as 

\begin{equation}
\hat{H}(t) = \bar{\xi} \int d k_3 dk_4 G(k_3,k_4,t) \hat{a}_3^\dagger (k_3) \hat{a}_4 (k_4)  + h.c.,
\label{eq:main}
\end{equation}
where all pump parameters and waveguide length are grouped together into  $\bar{\xi} : = \xi ( \Delta \phi, P_1, P_2, L,  \sigma_1, \sigma_2, R, \gamma)$, in which $\Delta \phi$ is the relative phase between both pumps, $L$ the length of the medium and $\gamma$ is the nonlinear coefficient of the TODFG process. The interacting fields will evolve within the medium obeying the evolution operator $\hat{U} \equiv \hat{U} (t_1,t_0) = \exp \{-i/\hbar \int_{t_0}^{t_1} dt' \hat{H} (t') \}$, where after integration from $t_0 \to - \infty$ to $t_1 \to \infty$, which applies given that we are interested in the fields long after the interaction medium \cite{Helt2015} and ignoring time-order effects \cite{Brecht2011}, can be expressed as follows

\begin{equation}
	\begin{aligned}
		\hat{U}  = & \exp \Big\{ -i( 2\pi  \bar{\xi} /(\mathcal{N}\hbar))\int d k_3 dk_4 G(k_3,k_4) \hat{a}_3^\dagger (k_3)  \hat{a}_4 (k_4) \\ & + h.c.,
		\Big\},
	\end{aligned}
	\label{eq:U0}
\end{equation}

\noindent where $\mathcal{N}$ is a normalization constant such that $\int dk_3 dk_4 |G(k_3,k_4)|^2 =1$, and $\hbar$ is the reduced Planck's constant. The coupling function $G(k_3,k_4)$, will be hereafter called frequency-conversion map (FCM) between modes 3 and 4. It is proportional to the probability amplitude of converting a photon with wave-vector $k_4$ to another one with $k_3$. Without loss of generality, photons in modes 3 and 4 are considered as lying in very different spectral regions, so as they commute for the sake of simplicity. The FCM can be written in its Schmidt decomposition \cite{URen2005,Brecht2011} as follows
\begin{equation}
	G(k_3, k_4 ) = \sum_j \sqrt{\kappa_j} \varphi_j(k_3)\phi_j^{*} (k_4),
	\label{eq:Schmidt}
\end{equation}
therefore the FCM is separable in its TMs. It is of particular relevance to write the evolution operator from Eq. (\ref{eq:U0}) with substitution of (\ref{eq:Schmidt}) in a similar fashion as in ref. \cite{Brecht2011}, in the simpler following form

\begin{equation}
	\hat{U} := \hat{U} (-\infty, \infty) = \exp \left\{ -i  \sum_j  \left( \theta_j \hat{A}_j^\dagger \hat{B}_j + \theta_j^*\hat{A}_j\hat{B}_j^\dagger \right)  \right\},
	\label{eq:U}
\end{equation}
where $\theta_j =2\pi \bar{\xi}\sqrt{\kappa_j} /( \mathcal{N} \hbar)$,  $\hat{A}_j^\dagger = \int dk_3 \varphi_j (k_3) \hat{a}_3^\dagger (k_3)$ and $\hat{B}_j = \int dk_4 \phi_j^{*} (k_4)  \hat{a}_4 (k_4)$. It is assumed that $\hat{A}$ modes commute with $\hat{B}$ modes, i.e. $\sum_{i,j}[\hat{A}_i,\hat{B}_j^\dagger] = 0$. Based on previous definitions, we proceed to define  an \textit{ad hoc} computational basis as follows 
\begin{equation}
\begin{aligned}
\vert 0 \rangle^C :&= \int dk_3 f(k_3) \hat{a}_3^\dagger (k_3) \vert \text{vac} \rangle = \sum_j \lambda_j A_j^\dagger \vac,\\   \vert 1 \rangle^C :&= \int dk_4 f(k_4) \hat{a}_4^\dagger (k_4) \vert \text{vac} \rangle = \sum_j \lambda_j \hat{B}^\dagger_j \vac , 
\end{aligned}
\label{eq:compBasis}
\end{equation}
where $\ket{0}^C$ and $\ket{1}^C$ are single photon states with spectral amplitude given by $f(k_i)$, normalized as $\int dk_i \vert f(k_i) \vert ^2 = 1$, and the expansion coefficients in the Shcmidt basis as $\lambda_j = \int dk \upsilon^*_j (k) f(k)$, with $\upsilon \in \{ \phi, \varphi \}$. Note that under these assumptions,  the elements of the computational basis are equivalent to the path-basis \cite{Monteiro2015}, with the only difference of being distinguished by energy instead that by their spatial localization. By substitution of Eq. (\ref{eq:Schmidt}) into (\ref{eq:U0}), the evolution of any linear combination with the states in Eq. (\ref{eq:compBasis}) will be governed by the Schmidt coefficients $\kappa_j$ and will be strongly related to the overlap of each TM-mode with the elements of the computational basis. It is therefore relevant, to determine de evolution of input operators to output operators as follows  

\begin{equation}
	\hat{A}_j^\dagger  \xrightarrow{\hat{U}} \cos ( \theta_j ) \hat{A}_j^\dagger - i e^{i\phi } \sin (\theta_j) \hat{B}_j ^\dagger,
\end{equation} 
\begin{equation}
	\hat{B}_j^\dagger  \xrightarrow{\hat{U}}   - i e^{-i\phi } \sin (\theta_j) \hat{A}_j ^\dagger + \cos ( \theta_j ) \hat{B}_j^\dagger,
\end{equation}
which resembles the BS transformation, as has been previously noted \cite{Brecht2015,Brecht2011}. With this, it is possible to define a \textit{single-qubit logic} (SQL), where the Hamiltonian can be written as a sum of interaction pairs between the $j$-th TM modes, where we write $\ket{\phi_j} = \ket{\phi}_j \otimes \ket{j}$ and  $\ket{\varphi_j} = \ket{\varphi}_j \otimes \ket{j}$. Note that the states $\{\ket{\varphi}_j, \ket{\phi}_j\}$ lay in a two dimensional basis, while $\{\ket{\varphi_j}, \ket{\phi_j}\}$ spans the complete space. Previous considerations allows us to write the time integrated Hamiltonian for each TM pair  $\ket{\phi}_j$ and $\ket{\varphi}_j$ as
\begin{equation}
	\int dt \hat{H}(t) = \sum_j \left[\int dt \hat{H}_j (t) \right]\otimes \ket{j} \bra{j},
	\label{eq:Hmatrix0}
\end{equation}

\noindent where each interaction pair is as follows 

\begin{equation}
	\int dt \hat{H}_j (t)  = \frac{ 2 \pi \sqrt{\kappa_j}}{\mathcal{N}} \left(\bar{\xi}  \hat{A}_j^\dagger \hat{B}_j + \bar{\xi}^* \hat{A}_j \hat{B}_j^\dagger \right),
\end{equation}

\noindent which can be written in a matrix TM-pair basis as follows

\begin{equation}
	\int dt \hat{H}_j (t)  = \frac{2\pi}{\mathcal{N}} \begin{pmatrix}
		0 & \sqrt{\kappa_j} \bar{\xi}^* \\
		\sqrt{\kappa_j} \bar{\xi} & 0
	\end{pmatrix}	=\hbar \Theta_j e^{-i\phi}\hat{T}(2\phi)\hat{\sigma}_x,
\label{eq:Hmatrix}
\end{equation}

\noindent where $\Theta_j = |\theta_j|$, $\phi = \arg (\theta_j)$ and
$\hat{T}= \vert \varphi \rangle_j {}_j\langle \varphi \vert + e^{i\phi}\vert \phi \rangle_j {}_j\langle \phi \vert$. Note that $\phi$ only depends on the pumps' phase difference and is constant for all $j$. Under this representation, every $H_j$ is a $2\times2$ matrix, and the whole Hamiltonian is described by a $2n \times 2n$, for $n$ TM-pairs. Basically, in this representation we have $n$ qubit interactions, which could result more intuitive. In addition, in this proposal we aim at maximizing only one of those interactions while suppressing the others.



\section{Qubit preparation and rotation} \label{sec:qubpreprot}

The qubit preparation process requires to generate, in principle, a vector state capable of spanning the whole Bloch sphere for a given computational basis. We restrict our analysis to the basis in Eq. \eqref{eq:compBasis} and assume that a single photon in a pure state is available. This state will prepare a qubit after evolving within the medium. Also, the feasibility of heralded pure photon states has been widely discussed and demonstrated by the optics community \cite{Kaneda2016,Garay-Palmett2007,URen2005,Takeuchi2014,Cohen2009,Yamamoto2005,Pittman2005,Lounis2005}. Specifically, the initial state is taken as
\begin{equation}
\ket{\psi}_0 = \ket{0}^C,
\end{equation}
as defined in Eq. (\ref{eq:compBasis}), which is evolved by the evolution operator in Eq. (\ref{eq:U}) into the qubit state
\begin{equation}
	\begin{aligned}
 \ket{\psi} &= \hat{U} \ket{\psi}_0 \\
 & = \sum_j \lambda_j \left(  \cos \left( \theta_j \right) \hat{A}_j^\dagger - i e^{i\phi} \sin \left( \theta_j \right) \hat{B}_j^\dagger   \right)  \vac.
\end{aligned}
 \label{eq:psi1}
\end{equation}

The usefulness of Eq. (\ref{eq:psi1}) will depend on its similarity with the defined computational basis, i.e., that the states defined as $\ket{0}^C$ and $\ket{1}^C$ remain with the same spectral envelope as initially defined, which is unlikely, as the different coefficients $\kappa_j$ will promote selective frequency conversion of photons, changing the structure of their spectra. In summary, a qubit of the form $\ket{\psi_{qubit}} = \left( \cos \theta \sum_j \lambda_j \hat{A}^\dagger_j  -ie^{i\phi} \sin \theta \sum_j \lambda_j \hat{B}^\dagger_j \right) \vac$ can be prepared if and only if  $\theta_j$ was independent of the TM index, which is not the case. To overcome this drawback, we propose three different scenarios with distinct implications as follows:

\begin{enumerate}[i)]
	\item The computational basis is described by only one TM of the same Schmidt pair as in Eq. (\ref{eq:Schmidt}), i.e., there exists one $j$ for which $\lambda_j =1$ and $\lambda_{i\neq j} = 0$. In other words, the input state is already a qubit in the form
	\begin{equation}
		\ket{\psi_{\text{in}}^{\text{ideal}}} = x_j^{in} \ket{\phi_j} + y_j^{in} \ket{\varphi_j},
		\label{eq:qubitTM}
	\end{equation} 
	
	where the states $\ket{\varphi_j} = \hat{A}_j^\dagger \vac $ and $\ket{\phi_j} = \hat{B}_j^\dagger \vac$ are single TM states, and $x_j^{in (out)}$ and $y_j^{in(out)}$ are the corresponding coefficients of the linear combination of TMs at the input (output) of the medium. With an input state in this form, even if $\hat{U}$ is expanded on many Schmidt modes, the only allowed transformation is between the two modes of the computational basis. In this scenario, a qubit of TMs could easily be prepared from a pure photon state as follows
	\begin{equation}
	\begin{aligned}
		\ket{\psi}^{C_j} &= \hat{U} \hat{A}^\dagger_j \vac =  \left(x_j^{out} \hat{A}_j^\dagger + y_j^{out} \hat{B}_j^\dagger\right) \vac \\
		& =  x_j^{out} \ket{\phi}_j + y_j^{out} \ket{\varphi}_j,
		\end{aligned}
		\label{eq:met1}
	\end{equation}

	where the initial state given by $\hat{A}^\dagger_j \vac$ was taken as an example for simplicity. This method requires an initial state in a pure and specific TM and an FCM that couples the specific input TM to another one. Given the input state in only one TM perfectly overlapping with one TM of the FCM, the preparation gets unitary efficiency.

	
	\item The FCM shown in Eq. (\ref{eq:Schmidt}) contains only one Schmidt pair, and the operator in Eq. \eqref{eq:U} is 
	\begin{equation}
	\hat{U}_j = \exp \left\{ -i \left( \theta_j \hat{A}_j^\dagger \hat{B}_j + \theta_j^*\hat{A}_j\hat{B}_j^\dagger \right)  \right\},
	\label{eq:FCMIdeal}
	\end{equation}
	so as even if the input state is spanned by many TMs, only one is coupled and transformed by the FCM. Let the initial state be spanned as $\ket{\psi_0} = \sum_i \lambda_i \hat{A}_i^\dagger \vac$, for which $\hat{A}_j$ couples with $\hat{B}_j$ as in Eq. \eqref{eq:FCMIdeal}. In this scheme, the FCM itself will determine the useful computational basis. In this scheme, the output state is
	\begin{equation}
		\begin{aligned}
		\ket{\psi} = \hat{U}\ket{\psi_0} =\mathcal{N} \left( \ket{\psi}^{C_j} + \sum_{i\neq j} \lambda_i \hat{A}_i^\dagger \vac \right),
		\end{aligned}
		\label{eq:met2}
	\end{equation}
	where $\mathcal{N}$ is another normalization constant, and $\ket{\psi}^{C_j}$ is the same as in Eq. \eqref{eq:met1}. The evolved state shown in Eq. (\ref{eq:met2}) implies that the useful computational basis $\ket{\psi}^{C_j}$ is somehow hidden after the action of the FCM, and further discrimination is required to purify the output state and keep $\ket{\psi}^C$ only. 
	
	 \item Both the initial state and the FCM contain a single TM pair. The evolved state is equivalent to the one in Eq. (\ref{eq:met1}), with the main advantage that there is only one Schmidt pair in the interaction. This can be seen from the state in Eq. \eqref{eq:qubitTM} and the mapping function in Eq. \eqref{eq:FCMIdeal}. With this, we construct the ideal evolved state as follows
	 \begin{equation}
	 	\ket{\psi_{\text{out}}^{\text{ideal}} } = \hat{U}_j \ket{\psi_{\text{in}}^{\text{ideal}}} = x_j^{out} \ket{\phi_j} + y_j^{out} \ket{\varphi_j}.
	 	\label{eq:PsiOutIdeal}
	 \end{equation}
	 Note that Eqs. \eqref{eq:PsiOutIdeal} and \eqref{eq:met1} are equivalent. This shows the importance of the initial state when the FCM couples more than two modes. 
\end{enumerate}

We have called $\ket{\psi}^{C_j}$ the useful qubit in a sense that contains only a TM pair, and therefore, once a qubit is prepared can be transformed to other states by the action of a parameterized FCM. These transformations can be directly implemented as described in Eqs. \eqref{eq:Hmatrix0} and \eqref{eq:Hmatrix}, where the $\hat{\sigma}_X$ ($\hat{\sigma}_Y$) Pauli gate can be implemented up to a global phase for $\phi = 0$ ($\phi = \pi/2$), for the corresponding condition 
\begin{equation}
	\theta_j = 2\pi \bar{\xi} \sqrt{\kappa_j} / \mathcal{N} \hbar = (2n + 1) \pi/2,
	\label{eq:conditionGate}
\end{equation}
for integer $n$. The remaining $\hat{\sigma}_Z$ Pauli gate can be implemented by successive application of $\hat{\sigma}_X$ and $\hat{\sigma}_Y$.

 The realistic action of the FCM over Gaussian input states will be analyzed in the following section focused on engineered-dispersion waveguides so as no pulse shaping is required.

\subsection{Ideal transformation}

\begin{figure}
    \centering
	\includegraphics[width=0.48\textwidth]{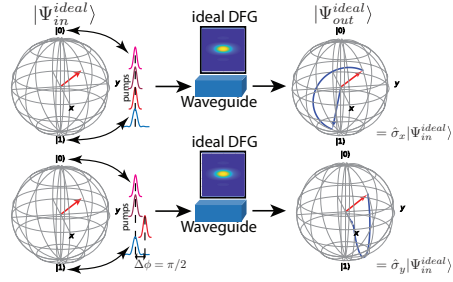}
	\caption{$\hat{\sigma_X}$ and $\hat{\sigma_Y}$ implementation by an ideal FCM transformation in a waveguide where only fumdamental TM modes $\phi_0$ and $\varphi_0$ are coupled, with an input qubit in a linear combination of them.}
	\label{Fig:IdealDFG}
\end{figure}

We will define an ideal transformation, as the one generated by an FCM with only one TM pair, which produces a transformation as the one in Eq. \eqref{eq:PsiOutIdeal}. To analyze the ideal situation we will go further and assume that both the FCM and the input interacting fields are restricted to the same TM pair. Let's assume a single photon in a linear combination of fundamental TM modes $\ket{\Psi_{in}^{ideal}}$, described by Eq. \eqref{eq:qubitTM}, enters into an ideal DFG waveguide, as shown in Fig.  \ref{Fig:IdealDFG}, this initial state evolves within the medium through a unitary transformation, for which, if the condition in Eq. \eqref{eq:conditionGate} is fulfilled, the $\hat{\sigma}_X$ and $\hat{\sigma}_Y$ are applied, as shown in the figure. Note that apart from the condition in Eq. \eqref{eq:conditionGate}, only a relative phase modulation $\Delta\phi$ is required to select one transformation over the other.

Perfect mappings, as the one shown in Fig. \ref{Fig:IdealDFG}, can be generated by tailoring the nonlinear interaction by shaping the pump pulses, which has already been proven experimentally \cite{Brecht2013}. Contrary to this, our proposal uses standard (Gaussian) shaped pulses while the interacting medium is engineered. 



\section{Frecuency map modeled by a third-order nonlinear waveguide} \label{sec:specificMAP}

The general transformation modeled by the FCM can be straightforwardly adapted to a single-mode nonlinear waveguide with co-polarized fields for which the parameters in Eqs. \eqref{eq:main} are as follows



\begin{subequations}
	\begin{equation}
		\bar{\xi} = \frac{3 \sqrt{2} \hbar L e^{i \Delta \phi} }{ 8\pi^{5/2}   }  \gamma   \sqrt{\frac{P^{av}_1 P^{av}_2 n_1(\omega_1^0) n_2 (\omega_2^0) }{R_1 R_2 \sigma_1 \sigma_2 \omega_1^0 \omega_2^0 }} 
	\end{equation}
	\begin{equation}
		\begin{aligned}
			\tilde{G}( \omega_3, \omega_4 )   & = \left(\frac{\omega_3 \omega_4}{n(\omega_3) n(\omega_4)} \right)^{1/2} \int d\omega_2 \sinc \left( \frac{L}{2} \Delta k \right) \\
			& \quad \times e^{-i L \Delta k /2} \alpha_2'(\omega_2 ) \alpha_1'(\omega_2 - \omega_3 + \omega_4),
		\end{aligned}
	\end{equation}
\end{subequations}


\noindent where the longitudinal propagation within the medium allows for a change from wave-vectors to frequencies, $\alpha_i'(\omega ) = \exp \left( -(\omega - \omega_i^0)^2 /\sigma_i^2 \right) $ are the Gaussian spectral envelopes of the pump pulses, centered at $\omega_i^0$ with bandwidth $\sigma_i$ and average power of $P_1^{av}$, $P_2^{av}$, repetition rates $R_1$, $R_2$ respectively, $L$ is the waveguide length, $e^{i\Delta \phi}$ takes into account the phase difference between both pumps, $\gamma$ is the nonlinear coefficient that takes into account the effective area of the medium for the interacting fields, and $\tilde{G}$ stands for a non-normalized FCM function. 

Variations on the geometrical characteristics of the waveguide will conduct to variations on the Schmidt decomposition, and therefore to a variation in the TM structure of the FCM, i.e., the terms $\kappa_j, \varphi_j$ and $\phi_j$ shown in Eq.  (\ref{eq:Schmidt}) will change. However, it is interesting to note that varying $e^{i\Delta \phi}\sqrt{P_1^{av} P_2^{av}}$ can potentially change the coupling between TM Pairs, while keeping the TM structure intact, assuming that pumps do not change in time. As can be seen from Eqs. \eqref{eq:U} and \eqref{eq:Hmatrix}, those variations are equivalent to changing the interaction time in the evolution. It is also possible to use group velocity dispersion matching techniques in a similar fashion as in \cite{Garay-Palmett2007}, so as the FCM is spanned by the minimum amount of Schmidt modes, or contains the desired modes.

In what follows, we will select a particular FCM that is close to a separable one and use this map to show how to create qubits and quantum gates by varying user-controllable parameters while Gaussian single-photon states are used as input modes. Also, the effect of non-separable FCM on Gaussian states will be analyzed.

\subsection{Non-ideal FCM with Gaussian input states}

\begin{figure}
    \centering
	\includegraphics[width=0.48\textwidth]{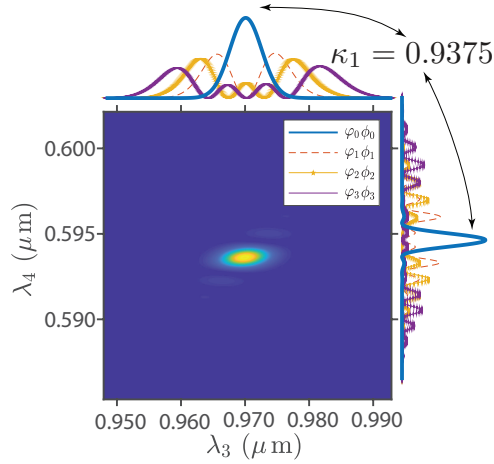}
	\caption{Non-ideal frequency conversion map, for which the first four TMs are shown}
	\label{Fig:NoIdealDFG}
\end{figure}

In general, for a realistic medium, a non ideal FCM must be considered. Here we model an experimental situation whose FCM is close to meet separability, but with a small contribution from different TM pairs. Specifically, we consider a ridge waveguide of Si${}_3$N${}_4$ on a SiO${}_2$ substrate, with a height $h=3 \mu m$ and width $w = 0.9 \mu m$ in the transverse coordinates. We also assume two pulsed pumps, where the first (second) one is centered at $\lambda_{p_1} = 1.55 \mu m$ ($\lambda_{p_2} = 0.77 \mu m$) with bandwidth $\sigma_1 = 0.3 \times 10^{12}$ rad/s ($\sigma _2 = 10 \times 10^{12}$ rad/s), average power of $P_1^{av} = 50 \mu$W ($P_2^{av} = 100 \mu$W) and the same repetition rate $R = 10$ MHz, with a length $L = 1$cm. Fig. \ref{Fig:NoIdealDFG} shows the corresponding FCM for the previous waveguide, where it can be seen that the pair of fundamental modes interact most. 

For this non-deal situation, we assume that a single-photon enters the medium in mode 3, because this mode is Gaussian-like, as can be seen from Fig. \ref{Fig:NoIdealDFG}, with the following normalized spectral amplitude
\begin{equation}
	f_{in} (\omega) = \left( \frac{2}{\pi \sigma_{in}^2} \right)^{1/4} e^{ \frac{-(\omega-\omega^0_{in})^2}{\sigma_{in}^2}}.
	\label{eq:spectral}
\end{equation}

Therefore, the input state to the medium is of the form
\begin{equation}
	\ket{\psi_{in}} = \int d\omega_3 f_{in} (\omega_3) \hat{a} ^\dagger_3 (\omega_3) \vac .
	\label{eq:PsiIn}
\end{equation}

The feasibility of having Gaussian single-photon states has already been discussed in references \cite{Aaronson2011,Heuck2019}. In general, the input state in Eq. \eqref{eq:PsiIn} will evolve to the following output state. 
\begin{equation}
	\begin{aligned}
	\ket{\psi_{out}^r} &= \hat{U}\ket{\psi_{in}} =\mathcal{N} \Big( \ket{\psi}^{C_0} + \sum_{j\neq 0} x_j^{out}  \hat{A}_j^\dagger \vac \\ 
	& \quad+ \sum_{j\neq 0} y_j^{out} \hat{B}_j^\dagger \vac  \Big),
\end{aligned}
\label{eq:RealOutput}
\end{equation}
where $\ket{\psi}^{C_0}$ represents the sought qubit state, while the coefficients $x_{j}^{out} (y_{j}^{out})$ represent the evolution of the modes $\hat{A}_i (\hat{B}_i)$ for $i\neq 0$ at the input that where coupled to the FCM and evolved accordingly into modes $\hat{A}_i$ or $\hat{B}_i$ at the output of the waveguide. It is also the presence of not null $\{x,y\}_{j}^{out} $ coefficients which decreases the proposed quantum gates fidelity.

Note that, by modulating the amplitude spectrum of the input photon, it could be possible to excite only the fundamental TM mode $\varphi_0$ and therefore, create a superposition at the output, like the one in Eq. \eqref{eq:met1}. However, we will assume the state in Eq. \eqref{eq:RealOutput} as the realistic output state from a given waveguide.


\subsection{Single-qubit logic}

In this work, we are interested in investigating the feasibility of the proposed scheme to perform one qubit preparations and transformations by starting with single-photon Gaussian wave-packets. To accomplish this, we rely on a waveguide geometry in such a way that the overlap between the input mode is maximized to the fundamental TM in the FCM. To do so, we define the following fidelity measure
\begin{equation}
	\mathcal{F} = |\bra{\psi_{out}^r} \psi\rangle^{C_0}|^2,
	\label{eq:Fid}
\end{equation}
that will quantify how close is the output state $\ket{\psi_{out}^{r}}$ to an ideally prepared qubit state $\ket{\psi}^{C_0}$, where for the former the FCM couples many TM pairs in general and the input state is also composed of a linear combination of more than a single TM, while the later only couples one TM pair and the input state contains one of those TMs. Note that we have defined the successful rate of qubit preparations as the fidelity measure, which is equivalent to arbitrary transformations given the linearity of the FCM action on the input state.
 
\begin{figure}
    \centering
	\includegraphics[width=0.48\textwidth]{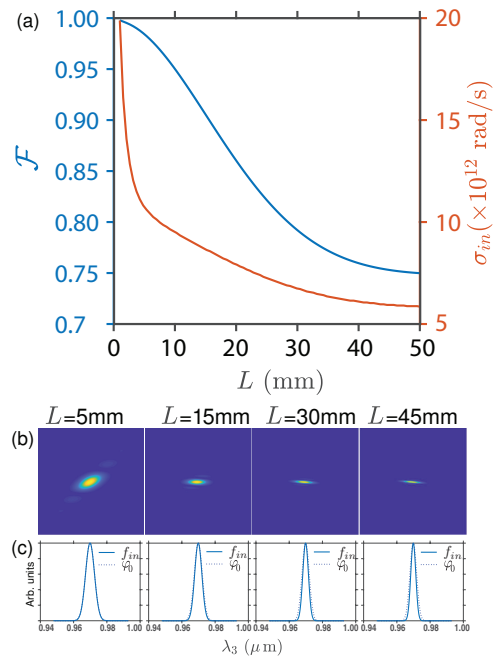}
	\caption{ (a) Single-qubit logic fidelity vs interaction length $L$ for optimal values of the Gaussian input bandwidth, (b) FCM for different interaction lengths with remaining parameters fixed, (c) $\varphi_0$ and the optimal Gaussian input state.}
	\label{Fig:Fid1}
\end{figure}

We now proceed to evaluate the fidelity defined in Eq. \eqref{eq:Fid} while some user-accessible parameters are varied. First, we maintain fixed all parameters except for the waveguide length $L$, which is shown in Fig. \ref{Fig:Fid1}(a), where the used strategy is as follows: for each $L$, the optimal bandwidth $\sigma_{in}$ (orange line) of an input state centered at the same wavelength that the first corresponding TM, $\varphi_0$ in this case, is selected, such that the fidelity is maximal (blue line). 

It is shown in Fig. \ref{Fig:Fid1}(a) that higher fidelities are obtained for shorter lengths; this can be explained by a better overlap of the Gaussian input function with the fundamental TM  $\varphi_0$. FCMs for different waveguide lengths with the corresponding spectra for $\varphi_0$ and the Gaussian input state are shown in Figs. \ref{Fig:Fid1}(b)-(c). A remarkable feature of this proposal, is that  a separable FCM is not needed to obtain high fidelities for input Gaussian input states. 

\begin{figure}
    \centering
	\includegraphics[width=0.38\textwidth]{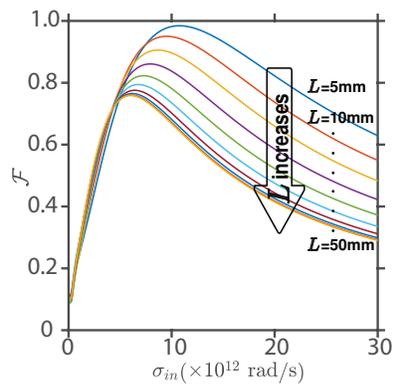}
	\caption{Single-qubit logic fidelity vs variations in the Gaussian input state bandwidth for different interaction lengths}
	\label{Fig:Fid2}
\end{figure}

Now we focus on a set of specific lengths ranging from 5mm to 50mm and analyze the fidelity vs. variations on the input state bandwidth, which is shown in Fig. \ref{Fig:Fid2}. The figure shows that high fidelities are attainable for a wide bandwidth range for each length, and that there exists an optimal length when Gaussian input states are considered. The figure also shows that while the center wavelength of the input state is held fixed.

\begin{figure}
    \centering
	\includegraphics[width=0.42\textwidth]{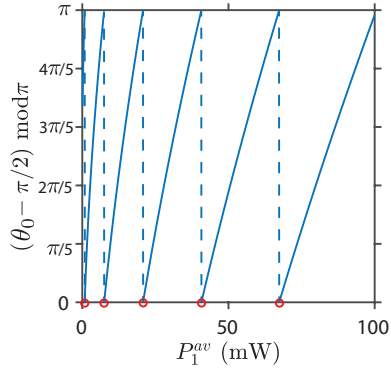}
	\caption{Deviations from $\hat{R}_{X(Y)}(n\pi) = e^{-i n \pi \hat{\sigma}_{X(Y)}/2}$ vs average power of pump 1 $P_1^{av}$. Red circles indicate different values of $P_1^{av}$ that will produce a Pauli gate.}
	\label{Fig:Fid3}
\end{figure}

Finally, we evaluate the fidelity of obtaining a Pauli gate action by the nonlinear interaction vs the average pump power 1, as shown in Fig. \ref{Fig:Fid3}, while the other parameters are kept fixed. This figure shows how far the rotation given by the interaction Hamiltonian to the input state is from the conditional $\theta_0$ in Eq. \eqref{eq:conditionGate}, for which the rotation fullfils $\hat{R}_{X(Y)}(n\pi) = e^{-i n \pi \hat{\sigma}_{X(Y)}/2}$ for this particular selection of parameters. The red circles show the average power $P_1^{av} $ that will produce the corresponding Pauli gate, which will finally depend on the phase difference $\phi$ between pump pulses.

\section{Conclusions}

We presented an alternative scheme to implement one qubit logic realizations assisted by means of third-order nonlinear interactions. In particular, qubit preparation was addressed by using Gaussian pulses as input states to a nonlinear waveguide, where the frequency conversion map of the difference frequency generation process served as the coupling between pairs of temporal modes with different energies, which were used to define the computational basis. One of the fundamental temporal modes from the frequency conversion map had a high overlap with a Gaussian pulse, which allowed for a high fidelity qubit preparation and transformations. Furthermore, a complete one qubit logic was presented. It was shown that varying some user-accessible parameters, one can select if a $\sigma_X$ or $\sigma_Y$ is applied. Fidelities for qubit preparation and Pauli gates action close to one were obtained for different conditions in the simulations. Even though, in this work, we were focused on single-photon states, it can be easily extended to other quantum states, like coherent or squeezed states. 

\section{Funding information}
		Consejo Nacional de Ciencia y Tecnología (Cátedras CONACYT 709/2018) and FORDECYT-PRONACES/194758/2020.
\section{Disclosures}
The authors declare no conflict of interest
	
\bibliography{CompsNL2020}







\end{document}